\documentclass[preprint,12pt, a4paper]{elsarticle}

\usepackage{amssymb}
\usepackage{hyperref}
\usepackage{tikz,float}

\journal{Software Impacts}

\begin{document}

\begin{frontmatter}



\title{tinygarden - A java package for testing properties of spanning trees}


 \author[1,2]{Manuel Dubinsky\corref{cor1}}
 \ead{mdubinsky@undav.edu.ar}
 
\author[3]{César Massri}

 \author[4]{Gabriel Taubin}
 
\cortext[cor1]{Corresponding author}

\address[1]{Ingeniería en Informática, Departamento de Tecnología y Administración, Universidad Nacional de Avellaneda, Argentina}
\address[2]{Departamento de Computación, Facultad de Ciencias Exactas y Naturales, Universidad de Buenos Aires, Argentina}
\address[3]{Departamento de Matemática, Universidad de CAECE, CABA, Argentina}
\address[4]{School of Engineering, Brown University, Providence, RI, USA}

\begin{abstract}
Spanning trees are fundamental objects in graph theory. The spanning tree set size of an arbitrary graph can be very large. This limitation discourages its analysis. However interesting patterns can emerge in small cases. In this article we introduce \emph{tinygarden}, a java package for validating hypothesis, testing properties and discovering patterns from the spanning tree set of an arbitrary graph.
\end{abstract}


\end{frontmatter}


\noindent
\section{Introduction}

Graph theory is a longstanding and well-established area of discrete mathematics. Graphs are abstract models of pairwise relations between entities in some domain. A graph $G=(V,E)$ is composed of $V$ -a set of \emph{vertices} (or \emph{nodes})- and $E$ -a set of \emph{edges}-. Each node is a point representing an entity and each edge is a line connecting two nodes.  For example we can model friendship relationships in a social network by associating a node to each person and connecting two nodes with an edge if the corresponding people are friends. 

\medskip

Let $G=(V,E)$ be the following graph
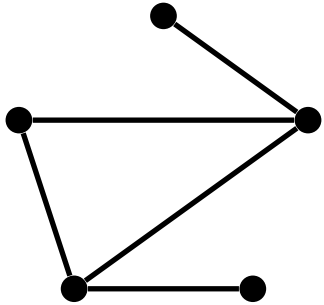
\begin{figure}[H]
\centering
\begin{tikzpicture}[shorten >=0pt,-]
  \tikzstyle{vertex}=[circle,fill=black,minimum size=10pt,inner sep=2pt]
  \tikzstyle{vertex_gray}=[circle,fill=gray,minimum size=10pt,inner sep=2pt]
  \node[vertex] (p1) at (234:2cm){};
  \node[vertex] (p2) at (162:2cm){};
  \node[vertex] (p3) at (90:2cm){};
  \node[vertex] (p4) at (18:2cm){};
  \node[vertex] (p5) at (-54:2cm){};
  \draw[color=black,line width=2pt] (p1) -- (p2);
  \draw[color=black,line width=2pt] (p3) -- (p4);
  \draw[color=black,line width=2pt] (p5) -- (p1);
  \draw[color=black,line width=2pt] (p1) -- (p4);
  \draw[color=black,line width=2pt] (p2) -- (p4);
\end{tikzpicture}
\caption{Example of a Graph}
\end{figure}
A \emph{subgraph} $G'=(V',E')$ is a graph such that $V' \subseteq V$,
$E'\subseteq E$
and such that $V'$ contains all endpoints of the edges in $E'$. 
A \emph{path} $p=(e_1, \dots, e_k)$  is a sequence of consecutive edges of $G$.
\begin{figure}[H]
\centering
\begin{tikzpicture}[shorten >=0pt,-]
  \tikzstyle{vertex}=[circle,fill=black,minimum size=10pt,inner sep=2pt]
  \tikzstyle{vertex_gray}=[circle,fill=gray,minimum size=10pt,inner sep=2pt]
  \node[vertex] (p1) at (234:2cm){};
  \node[vertex] (p2) at (162:2cm){};
  \node[vertex] (p3) at (90:2cm){};
  \node[vertex] (p4) at (18:2cm){};
  \node[vertex_gray] (p5) at (-54:2cm){};
  \draw[color=black,line width=2pt] (p1) -- (p2);
  \draw[color=black,line width=2pt] (p3) -- (p4);
  \draw[color=gray,dotted,line width=2pt] (p5) -- (p1);
  \draw[color=gray,dotted,line width=2pt] (p1) -- (p4);
  \draw[color=black,line width=2pt] (p2) -- (p4);
\end{tikzpicture}
\caption{Example of a Path}
\end{figure}
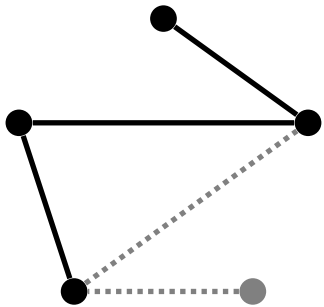
A \emph{cycle} in $G$ is a non-empty path in which the only repeated nodes are the first and last. 
\begin{figure}[H]
\centering
\begin{tikzpicture}[shorten >=0pt,-]
  \tikzstyle{vertex}=[circle,fill=black,minimum size=10pt,inner sep=2pt]
  \tikzstyle{vertex_gray}=[circle,fill=gray,minimum size=10pt,inner sep=2pt]
  \node[vertex] (p1) at (234:2cm){};
  \node[vertex] (p2) at (162:2cm){};
  \node[vertex_gray] (p3) at (90:2cm){};
  \node[vertex] (p4) at (18:2cm){};
  \node[vertex_gray] (p5) at (-54:2cm){};
  \draw[color=black,line width=2pt] (p1) -- (p2);
  \draw[color=gray,dotted,line width=2pt] (p3) -- (p4);
  \draw[color=gray,dotted,line width=2pt] (p5) -- (p1);
  \draw[color=black,line width=2pt] (p1) -- (p4);
  \draw[color=black,line width=2pt] (p2) -- (p4);
\end{tikzpicture}
\caption{Example of a Cycle}
\end{figure}
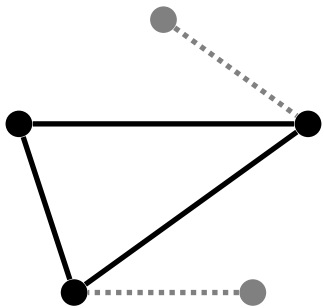
We say that G is \emph{connected} if there is at least one path between each pair of nodes. A \emph{tree} is a connected graph without cycles.  And finally, a \emph{spanning tree} $T \subseteq G$ is a tree that contains all the nodes of $G$.
\begin{figure}[H]
\centering
\begin{tikzpicture}[shorten >=0pt,-]
  \tikzstyle{vertex}=[circle,fill=black,minimum size=10pt,inner sep=2pt]
  \tikzstyle{vertex_gray}=[circle,fill=gray,minimum size=10pt,inner sep=2pt]
  \node[vertex] (p1) at (234:2cm){};
  \node[vertex] (p2) at (162:2cm){};
  \node[vertex] (p3) at (90:2cm){};
  \node[vertex] (p4) at (18:2cm){};
  \node[vertex] (p5) at (-54:2cm){};
  \draw[color=black,line width=2pt] (p1) -- (p2);
  \draw[color=black,line width=2pt] (p3) -- (p4);
  \draw[color=black,line width=2pt] (p5) -- (p1);
  \draw[color=gray,dotted,line width=2pt] (p1) -- (p4);
  \draw[color=black,line width=2pt] (p2) -- (p4);
\end{tikzpicture}
\caption{Example of a Spanning Tree}
\end{figure}
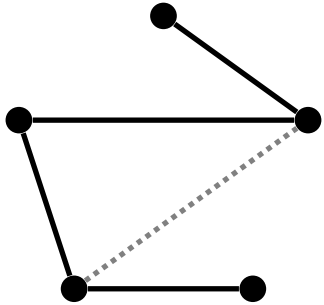

\medskip

Spanning trees are important in optimization \cite{wu2004spanning}, network design \cite{EPPSTEIN2000425}, VLSI interconnection \cite{Cong1996}, clustering \cite{Rokach2009}, complexity theory \cite{garey1979}, graph invariants \cite{Tutte1954}, fundamental cycle bases \cite{Kavitha:2009} and in many more areas of applied and theoretical sciences.

\medskip

The spanning tree set of an arbitrary graph can be very large \cite{Cayley1889}. This limit prevents the entire set from being explored. However, there are algorithms to generate all spanning trees of a graph (\cite{Read1975}, \cite{Gabow1978}, \cite{Matsui1993}) that are suitable in the case of small instances.

\medskip


\section{Software description}

The \emph{tinygarden} package is a self-contained group of \emph{java} classes that enables the exploration of the set of spanning trees of a given graph. It implements \emph{Matsui algorithm} \cite{Matsui1993} to generate it.

\medskip

The architecture is simple. The class \emph{SpanningTreesMatsui} organizes the whole process. Custom descendants of two class hierarchies: \emph{Collectors} and \emph{Processors} are invoked sequentially to process each spanning tree. \emph{Collectors} are useful for global analysis of the entire set, for example, count the number of spanning trees with a certain property. Processors, on the other hand, are useful for local tasks, such as generating a pretty-print version of a specific spanning tree.

\medskip

The package also implements helper classes, such as \emph{Graph}, \emph{SparseMatrix}, \emph{UnionFind}, etc. Since a \emph{Graph} can be built based on a text file containing its incidence matrix, it works well together with \emph{nauty} \cite{McKay:2014}, the well-known graph software.


\section{Limitations and improvements}
The scope of the \emph{tinygarden} package is restricted to specific graph theory problems. The most important limitation is the size of the instances that can be processed; it was conceived to explore the spanning tree set of all non-isomorphic graphs of at most 9 nodes. Consequently, two important improvements would be:

\begin{itemize}
	\item A distributed implementation of Matsui algorithm: in order to process larger graphs.
	\item Compute the spanning tree set size \cite{Kirchhoff1847}: to process graphs with smaller size compared to a prescribed threshold. 
\end{itemize}

The first version of the package was recently made public. At the moment is only being used by the authors.

\begin{figure}[t]
\includegraphics[scale=0.23]{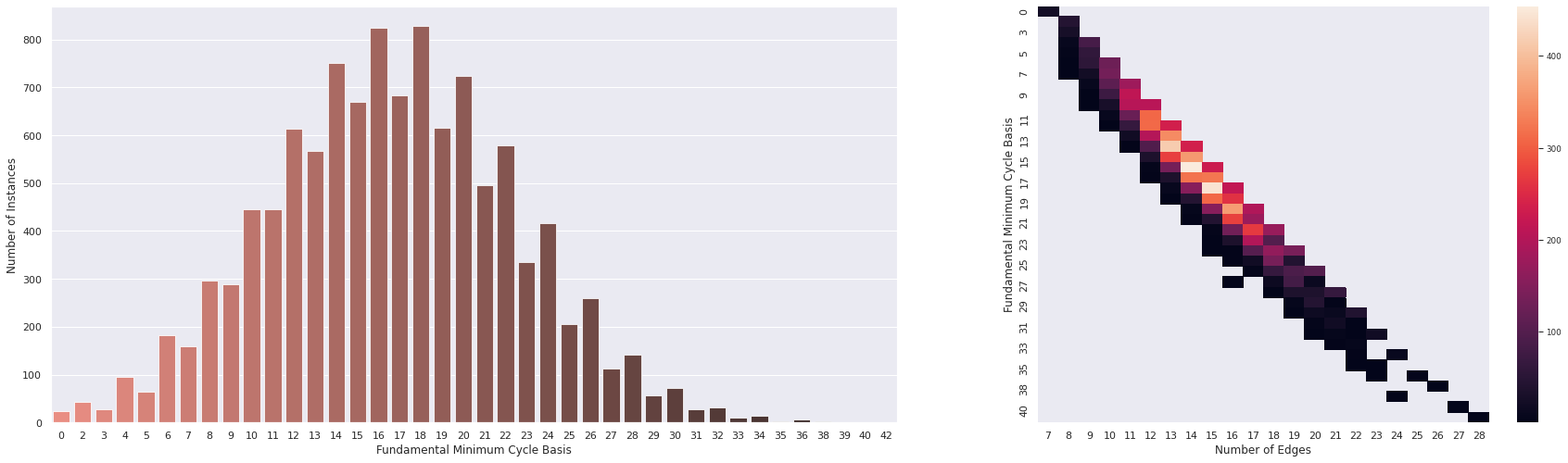}
\caption{Distribution of the Min FCB of the graphs of 8 nodes (left). Aggregated relation of the Min FCB with respect to the number of edges (right)}
  \label{fig:mfcb}
\end{figure}

\section{Impact}

The \emph{tinygarden} package is a tool for validating hypothesis, testing properties, and discovering patterns from the spanning tree set of an arbitrary graph. To the best of our knowledge this is the only tool with these features.

\medskip

In \cite{Dubinsky2021} we introduced the \emph{mstci problem}, the \emph{tinygarden} package proved to be effective in finding a pattern related to the \emph{intersection number} of complete graphs which led to its main result, and to set a solid basis for the conjecture posed in it. It was an important tool to evaluate a metaheuristic algorithm for graph integration \cite{Dubinsky2017}. We are currently using it to test an algorithm in order to find a good solution to the \emph{mstci problem}.

\medskip

The package can be used to analyze general questions, i.e.: list the spanning trees with minimum diameter of a graph, as well as to carry out statistical analysis, i.e.: calculate the distribution of the graphs of a fixed number of nodes with respect to the spanning tree with shortest total path length \cite{garey1979}.

\medskip

Based on our package, several NP-hard problems can be analyzed from a statistical perspective, for example Fig. \ref{fig:mfcb} shows the distribution of the \emph{Minimum Fundamental Cycle Basis} (or \emph{Min FCB})\cite{Deo:1982} of 8 node graphs. This information can be used to implement approximate or heuristic algorithms in order to find good solutions for these problems.

\medskip


\section*{Declaration of competing interest}

The authors declare that they have no known competing financial interests or personal relationships that could have appeared to influence the work reported in this paper.

\section*{Acknowledgements}

Prof. Massri was supported by Instituto de Investigaciones Matemáticas ``Luis A. Santaló'', UBA, CONICET, CABA, Argentina.

\medskip

Prof. Taubin was partially supported by the National Science Foundation grant number IIS-1717355.


\section*{References}





\bibliographystyle{elsarticle-num}
\bibliography{tiny_garden}


\section*{Current code version}
\label{}

\begin{table}[!h]
\begin{tabular}{|l|p{6.5cm}|p{6.5cm}|}
\hline
\textbf{Nr.} & \textbf{Code metadata description} & \textbf{Please fill in this column} \\
\hline
C1 & Current code version & 1.0 \\
\hline
C2 & Permanent link to code/repository used for this code version & \url{https://github.com/manudubinsky/tinygarden} \\
\hline
C3  & Permanent link to Reproducible Capsule & \url{https://codeocean.com/capsule/9539109/tree/v1} \\
\hline
C4 & Legal Code License   & MIT \\
\hline
C5 & Code versioning system used & git \\
\hline
C6 & Software code languages, tools, and services used & java \\
\hline
C7 & Compilation requirements, operating environments \& dependencies & jdk \\
\hline
C8 & If available Link to developer documentation/manual &  \url{https://github.com/manudubinsky/tinygarden/wiki} \\
\hline
C9 & Support email for questions & mdubinsky@undav.edu.ar \\
\hline
\end{tabular}
\caption{Code metadata (mandatory)}
\label{} 
\end{table}

\end{document}